\documentclass[a4paper]{IEEEtran}

\usepackage{graphicx}
\usepackage{latexsym}
\usepackage{amsfonts}
\usepackage{amssymb}
\usepackage{amsbsy}
\usepackage{color}
\usepackage{amsmath}
\usepackage{multicol}
\usepackage{float}
\usepackage{subfigure}
\usepackage{algorithm}
\usepackage{algorithmic}
\usepackage[dvips]{epsfig}

\newcommand{\RNum}[1]{\uppercase\expandafter{\romannumeral #1\relax}}

\begin{document}

\title{Sum-Rate Maximizing Cell Association \\via Dual-Connectivity}

\author{\IEEEauthorblockN{Minho Kim, Sang Yeob Jung, and Seong-Lyun Kim}\\
\IEEEauthorblockA{School of Electrical and Electronic Engineering, Yonsei University\\
50 Yonsei-Ro, Seodaemun-Gu, Seoul 120-749, Korea\\
Email: \{mhkim, syjung, slkim\}@ramo.yonsei.ac.kr}}

\maketitle

\begin{abstract}
This paper proposes a dual-connectivity (DC) profile
allocation algorithm, in which a central macro base station
(MBS) is underlaid with randomly scattered small base stations
(SBSs), operating on different carrier frequencies. We
introduce two dual-connectivity profiles and the differences
among them. We utilize the characteristics of
dual-connectivity profiles and their applying scenarios to
reduce feasible combination set to consider. Algorithm
analysis and numerical results verify that our proposed
algorithm achieve the optimal algorithm's performance within
5\% gap with quite low complexity up to $10^{-6}$ times.

\end{abstract}


\section{Introduction}

Due to spatio-temporal traffic variations coupled with the
continuously increasing demand for higher data rates, the
heterogeneity is regarded as a key characteristic of the
evolving fourth generation (4G) cellular wireless networks.
Deviating from traditional homogeneous macrocellular networks,
the 4G networks are multi-tier heterogeneous cellular networks
(HCNs) comprising existing macro base stations (MBSs) overlaid
with a diverse set of small base stations (SBSs) such as
picocells and femtocells \cite{green_small}--\cite{pimrc}.
Such deployment of SBSs inside the wide coverage area of MBSs
enables to provide high capacity and coverage blanket with low
power and low cost, resulting in an efficient and
cost-effective approach to cater for the expected demand for
data \cite{Hetnet}.


Despite its immense techno-ecocnomic benefits, however, the
{\it handover} issue should be addressed to support the
wide-scale deployment of SBSs \cite{Hetnet, Trends_small}. 
For example, the higher the density of SBSs, the more handovers
between user equipments (UEs) and their serving SBSs will be
triggered. 
This results in (1) extra traffic latency; (2)
increased signaling overhead; 
\begin{figure}[!hb]
\centering
\includegraphics[angle=0, width=3.3in]{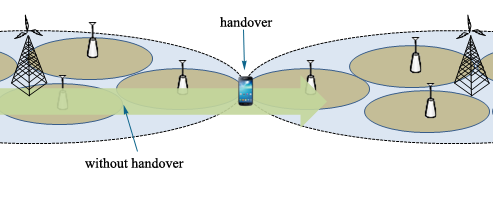}
\caption{The concept of Dual-Connectivity.}
\label{fig:handover}
\end{figure}
and (3) higher risk in radio
link failure (RLF), 
especially in high mobility scenario
\cite{Hetnet, Trends_small}. 
As a remedy, the dual
connectivity (DC) stemming from the control plane/user-plane
(C/U-plane) split architecture has been actively standardized
by the 3rd Generation Partnership Project (3GPP) as part of
the Release 12 specifications [see \cite{TR
36.842}--\cite{workitem} and the references therein]. As
illustrated in Fig. 1, the DC enables a user equipment (UE) to
maintain the connection to the MBSs and receive the C-plane
data (i.e., signaling messages) from them at all times. The
UEs in DC mode needs not to initiate the handover procedure
unless moving the coverage of other MBSs, thereby handling the
handover issue efficiently.

\begin{figure}[t]
\centering
   \subfigure[1A profile]{\centering
     \includegraphics[width=1.66in]{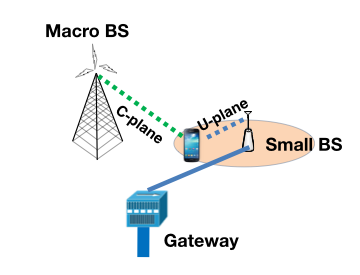} }
   \subfigure[3C profile]{
     \includegraphics[width=1.66in]{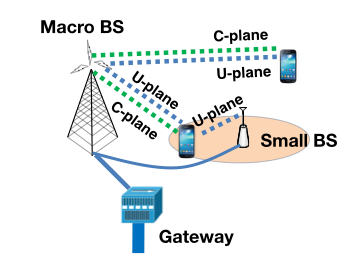}}
\caption{Two major types of Dual-Connectivity profiles.}
\label{fig:profiles}
\end{figure}

Even though DC is a promising technology to improve the
mobility performance, limited attention has been paid to the
capacity maximization, which is another important issue not to
be overlooked. There are two major types of DC profiles under
consideration by 3GPP: (i)1-A profile; and (ii) 3-C profile.
As depicted in Fig. 2, the UEs receive the U-plane data only
from the SBSs in the 1-A profile. In the 3-C profile, on the
other hand, the UEs receive the U-plane data from the MBSs and
the SBSs simultaneously \footnote{Another difference between
the 1-A profile and 3-C profile is the traffic flow in a wired
network from a Gateway to the MBSs (or SBSs). But we only
focus on the wireless traffic flow, throughout the paper.}. As
one of the solutions for the capacity maximization, the
maximum received power-based profile allocation scheme can be
exploited, but this may suffer from undesirable scheduling
starvation, i.e., the inability to guarantee the
quality-of-service for the most UEs. Achieving the capacity
maximizing solution while considering the issue of the
scheduling starvation is not trivial, due to the multiple
number of cases to be allocated, which makes the problem to be
NP-hard.


Motivated by the above discussions, in this paper, we propose
a suboptimal algorithm, which is more efficient but is
close to the optimum capacity. The remainder of this paper is
organized as follows: Section \RNum{2} describes the system
model. Section \RNum{3} formulates the capacity maximization
problem. Section \RNum{4} presents our proposed algorithm.
Section \RNum{5} gives numerical results to validate the
proposed studies, followed by concluding remarks in Section
\RNum{6}.

\section{System Model}

Consider the downlink of a two-tier HCN consisting of one
central MBS serving an area $\mathcal{A}$, overlaid with a set
$\mathcal{I}=\{1,2,\cdots, I\}$ of SBSs. We assume that each
SBS $i\in\mathcal{I}$ is uniformly distributed in
$\mathcal{A}$. Among potential deployment scenarios in DC, we
focus on the dedicated-channel deployment scenario in which
the MBS and the SBSs operate on different carrier frequencies.
Notice that this scenario is prioritized to support the DC in
Release 12 in terms of mobility robustness and UE throughput
enhancement [5-6]. We define We assume that a set $\mathcal{K}
= \{1, 2, \cdots, K\}$ of UEs are uniformly and independently
distributed in $\mathcal{A}$.

\subsection{Channel Model and UE Association Schemes}

Assume that $P_{m}$ and $P_{s}$ are the transmit powers from the MBS and the SBSs.
Over the time-period of interest, all the channel gains are fixed, i.e.,
the channels remain constant for each time-period, but possibly change over different time-periods.
The channel gain between UE $k$ and SBS $i$ is denoted by $h_{k,i}$, and
the channel gain between UE $k$ and the MBS is denoted by $g_{k}$.

Without loss of generality, every UE can be associated with the two-tier HCN simultaneously.
In DC, all the UEs receive the C-plane traffic from the MBS for handover robustness.
In this context, they should be always associated with the MBS.
Assume that UE $k\in\mathcal{K}$ is associated with the MBS.
Then, the resulting signal-to-noise ratio (SNR) for UE $k$ can be expressed as
\begin{equation}
\text{SNR}_{k,m} = \frac{P_{m}g_{k}}{w_{m}},
\end{equation}
where $w_{m}=B_{m}n_{m}$ with $n_{m}$ being the power spectral density of the background noise for the MBS.

For the association of SBSs,
each UE $k\in\mathcal{K}$ is associated with SBS $i\in\mathcal{I}$ that provides the maximum received power,
i.e., $i=\arg\max\left(P_{s}h_{k,i}: k\in\mathcal{K}, i\in\mathcal{I}\right)$.
Then, the corresponding single-to-interference-plus-noise ratio (SINR) for UE $k$ is modeled as
\begin{equation}
\text{SINR}_{k,i} = \frac{P_{s}h_{k,i}}{\sum_{j\in\mathcal{I}\backslash k}P_{s}h_{k,j} + w_{s}},
\end{equation}
where $w_{s}=B_{m}n_{s}$ with $n_{s}$ denoting the power spectral density of the background noise for the SBSs.

In practice, the UEs transmit channel quality information or
channel status information to the MBS, the MBS can estimate
the SINR (or SNR) of every UE [9-12]. Thus, we consider that
the MBS can assemble the SINRs (or SNRs) of the UEs in the
two-tier HCN, throughput the paper.

\subsection {Loading Model and Scheduling}

For each UE $k\in\mathcal{K}$,
the MBS needs to determine one of the DC profiles for capacity maximization.
In 1-A profile, UE $k$ receives the U-plane traffic only from the SBSs.
In 3-C profile, on the other hand, UE $k$ can receive the U-plane traffic from both the MBS and the SBSs.
To be more specific, the MBS can forward some of the U-plane traffic to the SBSs to relay it for UE $k$.

To this end, we assume that the MBS has a Data-Flow-Entity (DFE) for each UE $k$.
Mathematically, this can be expressed as
\begin{eqnarray}
D_{k} = (D_{k,m}, D_{k,i}), \forall k\in\mathcal{K}, \forall i\in\mathcal{I}
\end{eqnarray}
where $D_{k,m}, D_{k,i}\in\{0,1\}$,
and $m$ and $i$ denote the MBS and SBS $i$ that provides the maximum received power to UE $k$, respectively.
Note that $D_{k}=(0,1)$ represents the 1-A profile is allocated to UE $k$, while $D_{k}=(1,0)$ or $D_{k}=(1,1)$ for the 3-C profile.

Under the above framework,
we consider a round-robin scheduling mechanism for resource allocation.
In this case, we assume that all the UEs are always in service by at least one of the available tiers.
For the purpose of exposition,
we define the set of UEs receiving the U-plane traffic from the MBS by $\mathcal{K}_{m}=\left\{k|D_{k,m}=1\right\}\subset\mathcal{K}$,
and the set of UEs receiving the U-plane traffic SBS $i$ by $\mathcal{K}_{i}=\left\{k|D_{k,i}=1\right\}\subset\mathcal{K}$.
Then, the rate of UE $k\in\mathcal{K}_{m}$ can be written as
\begin{equation}
R_{k,m} =
\frac{B_{m}}{\sum_{k\in\mathcal{K}_{m}}D_{k,m}}\log\left(1+SNR_{k,m}\right),
\label{Eq:rate_m}
\end{equation}
where the MBS allocates the frequency bandwidth $B_{m}$ evenly to the UE $k\in\mathcal{K}_{m}$.
Similarly, the rate of UE $k\in\mathcal{K}_{i}$ is
\begin{equation}
R_{k,i} =
\frac{B_{s}}{\sum_{k\in\mathcal{K}_{i}}D_{k,i}}\log\left(1+SINR_{k,i}\right).
\label{Eq:rate_s}
\end{equation}

\section{Problem Formulation}

In this section, we now consider the DC profile allocation
(DCPA) problem over the two-tier HCN. To this end, the
optimization problem for capacity maximization can be
formulated as follows:
\begin{equation}
{\bf{DCPA}}:\max_{D_{k,m}, D_{k,i}}
\sum_{k\in\mathcal{K}_{m}}R_{k,m} +
\sum_{i\in\mathcal{I}}\sum_{k\in\mathcal{K}_{i}}R_{k,i},\quad\quad\quad
\label{Eq:MaxProb}
\end{equation}
\begin{equation}
\text{s.t.}\quad\;\; D_{k,m}\in\{0,1\}, \forall
k\in\mathcal{K},
\end{equation}
\begin{equation}
\;\quad\quad\quad\quad\quad D_{k,i}\in\{0,1\}, \forall
k\in\mathcal{K}, \forall i\in\mathcal{I},
\end{equation}
where $\mathcal{K}_{m}=\left\{k|D_{k,m}=1\right\}$, and
$\mathcal{K}_{i}=\left\{k|D_{k,i}=1\right\}$.

Note that this problem is not straightforward to solve,
since it is the integer programming problem (i.e., all of the decision variables take on the value 0 or 1).
This indicates that the problem is intractable and thus heuristic methods should be exploited instead.
In the following section, we will study how to design a sub-optimal algorithm.

\section{Proposed Algorithm}

One of the optimal solutions for Problem {\bf DCPA} is the
brute-force algorithm that takes into account all possible
combinations of data flow, $D_k$. It is hard to solve Problem {\bf DCPA}
with the brute-force algorithm since it has an extremely large number of cases to consider. In case of $K$
UEs in the network, the number of possible combinations to check is 3$^{K}$ since there three options for each users:

{\small{
\[
D_k  = ( D_{k,m} ,D_{k,i} )  = ( \,1,\,1\,) \,or\,( \,1,\,0\,)
\,or\,( \,0,\,1\,).
\]
\normalsize}\normalsize}

We devise a sub-optimal algorithm to reduce the complexity of
brute-force algorithm. In the first step, the proposed
algorithm creates a primitive matrix by $K$ rows by $I+1$
columns, and set a UE identity $k$ at position $ (k, i)$ of
the primitive matrix if $D_{k,i}$ equals one. If $D_{k,m}$
equals one, set a UE identity $k$ at position ($k$, $I+1$) of
the matrix. Then convert the primitive matrix to an initial
matrix by sorting its columns in descending order.

To simplify the notation, we identify $m$ with $I+1$ in this
section. Hence the largest $i$ connote the identity of an MBS
e.g. $D_{k,m} = D_{k,I+1}$, and BS $i$ is an MBS for $i = I+1$
or an SBS for $i < I+1$. The initial matrix implies that the
DC profile of each UE in a network is 3C type that both an MBS
and one of SBSs serve a UE simultaneously. The proposed
algorithm change the individual DC profiles of some UEs in the
initial matrix for maximizing network capacity, to either 1A
or another 3C at which only an MBS serves downlink
transmission to the UE.

\vskip 10pt \noindent {\bf Proposition 1} (\textbf{Necessary
condition to be an optimal solution}) {\it To maximize the
total network capacity in \eqref{Eq:MaxProb}, the UE groups that
each BS $i$ serves include a UE which have the highest SINR
toward the BS $i$.} \vskip 3pt

 {\it Proof }\ : Let $G_{opt,i}$ denote the group of UEs that $BS_i$ serves in an optimal
 solution, $G_{opt,i}  = \{ U_{1,i} ,\,U_{2,i} ,\, \cdots \,,\,U_{n,i} \}$
 and $\text{SINR}_{k,i}$ $>$ $\text{SINR}_{k+1,i}$ for $k \in \mathcal{K}_{i}$, where $\text{SINR}_{k,i}$ denotes SINR value at ${U_{k,i}}$
toward BS $i$. $U_{ \bar{k} ,i}$ denote the UE which is placed
at $(1,i)$ in initial matrix. We show that $U_{1,i} =
U_{\bar{k} ,i}$ by reductio ad absurdum.

 We assume that $U_{1,i} \ne U_{\bar{k},i}$. Then the UE
 corresponding to $U_{\bar{k},i}$ is placed in another
 column $i' \ne i$ because at least one BS serve each UE.
 There exists another valid combination which is equal to
 the optimal solution except for $G_{opt',i}$ = $G_{opt,i}$ $\cup$ $\{U_{\bar{k},i}\}$.
 Total throughput of BS $i$ in the optimal solution and the alternative solution,
 can be described as follows respectively:

{\small{\medmuskip=-2mu\thinmuskip=-2mu\thickmuskip=-2mu
\begin{eqnarray}\label{eq:opt}
\textsf{optimal solution: }& & \sum\limits_{k \in
\mathcal{K}_i}{\left[ R_{k,i} \right]} = \sum\limits_{k \in
\mathcal{K}_i}{\left[ {\frac{{B_{i} }}{{n}} \cdot \log (1 +
\text{SINR}_{k,i} )} \right]}
\end{eqnarray}
\normalsize}\normalsize}

{\small{\medmuskip=-2mu\thinmuskip=-2mu\thickmuskip=-2mu
\begin{eqnarray}\label{eq:alt}
\textsf{alternative solution:}& &\sum\limits_{k \in
\mathcal{K}_{i} \cup \{ 1 \} }{\left[ R_{k,i}
\right]}\nonumber
\\
&=&\sum\limits_{k \in \mathcal{K}_{i} \cup \{ 1 \} } {\left[
{\frac{{B_{i} }}{{n+1}} \cdot \log (1 + \text{SINR}_{k,i} )}
\right]}
%
%
%
%
%
\end{eqnarray}
\normalsize}\normalsize}

Subtracting \eqref{eq:opt} from \eqref{eq:alt} yields

{\small{\medmuskip=-2mu\thinmuskip=-2mu\thickmuskip=-2mu
\begin{eqnarray}\label{eq:b_result}
\frac{B_{i}}{{n + 1}}\log (1 +
\text{SINR}_{\bar{k},i})\nonumber
\\- \left[ \sum\limits_{k = 1}^{n} {\frac{B_{i}}{{n(n +
1)}}\log (1 + \text{SINR}_{k,i})} \right]
\end{eqnarray}
\normalsize}\normalsize}

Substituting $n/(n+1)n$ for $1/(n+1)$ in \eqref{eq:b_result}
yields

{\small{\medmuskip=-2mu\thinmuskip=-2mu\thickmuskip=-2mu
\begin{eqnarray}
= \ \ B_{i} \sum\limits_{k = 1}^{n} {\left[ {\frac{1}{{n(n +
1)}}\log \left( {\frac{{1 + \text{SINR}_{\bar{k},i} }}{{1 +
\text{SINR}_{k,i})}}} \right)} \right]}\label{eq:pos_result}
\end{eqnarray}
\normalsize}\normalsize}

\eqref{eq:pos_result} is positive because
$\text{SINR}_{\bar{k},i}> \text{SINR}_{k,i}$ for $k \in
\mathcal{K}_i$. Then, the sum of total throughput of each BS
in the alternative solution is greater than the optimal
solution's. Thus, $U_{1,i} = U_{\bar{k},i}$ in an optimal
solution. \hfill $\blacksquare$

\begin{algorithm}
\caption{Proposed Algorithm, main()} \label{main}
\begin{algorithmic}[1]

\STATE $servedset \gets \emptyset$

\FORALL{$i$}

    \STATE $currentset_{i} \gets \emptyset$

    \STATE add $U_{1,i}$ to the $currentset_{i}$

    \STATE add $currentset_{i}$ to the $servedset$

    \STATE $current_i \gets 1$

\ENDFOR

\FORALL{$i$}

    \STATE move()

    \STATE search()

\ENDFOR

\IF{all UEs are in $servedset$}

    \STATE terminate.

\ELSE

    \STATE go to step 7.

\ENDIF

\end{algorithmic}
\end{algorithm}

\begin{algorithm}
\caption{Proposed Algorithm, move()} \label{move}
\begin{algorithmic} [1]

\STATE $next_i \gets current_i$

\WHILE{$UE$ at ${next_i} \in servedset$}

    \STATE increase $next_i$ by 1

    \IF{$next_i$ is the end of column $i$}

        \STATE terminate.

    \ENDIF

\ENDWHILE

\end{algorithmic}
\end{algorithm}

\begin{algorithm}
\caption{Proposed Algorithm, search()} \label{search}
\begin{algorithmic}[1]

\FORALL{$i$}

    \STATE $comb \gets$ power set of $\{ U_{k',i} | current_i \le k' \le next_i \}$

    \STATE $bef_i \gets$ calculate sum-rate at BS$i$ serving UEs $\in currentset_i$

    \FORALL{$c \in comb$}

        \STATE $aft_{c,i} \gets$ calculate sum-rate at BS$i$ serving UEs $\in \{ currentset_i \, \cup \,
        c$ \}

    \ENDFOR

\ENDFOR

\STATE $best_{c,i} = \mathop {\arg}\limits_c \,\min (\,bef_i -
aft_{c,i} )$

\STATE $currentset_i \gets currentset_i \cup best_{c,i}$

\STATE $current_i \gets$ the largest row index in
$currentset_i$

\STATE $servedset \gets servedset \cup best_{c,i}$

%

\end{algorithmic}
\end{algorithm}

A maximum throughput of the network can be achieved when only top
SINR UEs in each BS area are in service. The basic concept
of the proposed algorithm is that a suitable UE
that minimizes the throughput degradation is admitted, until all the UEs are in service by at least one BS. 
Using Proposition 1, the proposed algorithm is summarized above.

\vskip 10pt \noindent {\bf Worst-Case Complexity}\\ The worst
case complexity of the proposed algorithm is $2^{K-1} -
2^{N-1} + (K-N)(N-1)$, where $K$ and $N$ denote the numbers of
UEs and BSs (including both MBSs and SBSs) respectively. The
worst case complexity can be obtained by considering the
situation when adding UEs to the $servedset$ which are
selected only in SBSs, because it makes a large gap between
$current_{macro}$ and $next_{macro}$ at \textbf{search()}. The
number of all possible combinations at \textbf{search()} is
$2$ \^{} $(next_{macro}-current_{macro})$ and it becomes the
dominant term in complexity so that the complexity situated
above is formulated as follows,


{\small{\medmuskip=-2mu

\begin{eqnarray}
&&\underbrace {\overbrace {2^{N-1} }^{macro} + \overbrace {N -
1}^{small}}_{1^\mathrm{st}\,\mathrm{iteration}} + \underbrace
{2^N + N - 1}_{2^\mathrm{nd}\,iteration} +  \cdots  +
\underbrace {2^{K
- 2} + N - 1}_{(K - N)^\mathrm{th}\,iteration}\nonumber\\
 &=& 2^{K - 1}
- 2^{N - 1} + (K - N)(N - 1)
\end{eqnarray}
\normalsize}\normalsize}


\noindent whereas the complexity of an optimal algorithm but
brute-force algorithm is $K \cdot 3^K$ since each combination
needs $K$ number of rate calculations.

\section{Performance Evaluation}

We show the result of the proposed sub-optimal algorithm's
performance by comparing with the full-search optimal
algorithm. The result includes comparisons with not only the
optimal solution, but also other conventional algorithms such
as assigning either 1A or 3C profile to all UEs and assigning
a UE to the only one closest BS. The specific parameters used
in our simulation are summarized in Table
\ref{tab:parameters}. Note that the power spectral density
$n_m$ for an MBS is higher than $n_s$. It is to reflect
interference coming from other MBSs which could exist if
there were multiple MBSs. In `3C only' algorithm, all UEs in a
network are in service by both BSs (MBS and SBS)\footnote{We
only consider 3C that assign both an MBS and an SBS for
downlink transmission even if it is possible to assign an MBS
only for downlink transmission.}, whereas in `1A only'
algorithm, all UEs are in service by an SBS only. In
`Stronger' algorithm, a UE is in service by more closer BS
between the MBS and the SBSs. All results in this section are
outcomes under four SBSs with a single MBS. Each result is
based on Monte Carlo simulation routines.

\begin {table}
\caption {Simulation Parameters} \label{tab:parameters}
\begin{center}
\begin{tabular}{|c||c|}

\hline
Network Size & 500m $\times$ 500m\\

$\alpha_{macro}$ & 4.5\\

$\alpha_{small}$ & 5\\

Macro Cell Bandwidth & 10 MHz\\

Small Cell Bandwidth & 10 MHz\\

Macro TX Power & 46 dBm\\

Small TX Power & 20 dBm\\

$n_m$ for Macro Cell & -90 dBm/Hz\\

$n_s$ for Small Cell & -140 dBm/Hz\\
\hline

\end{tabular}

\end{center}

\end {table}

\begin{figure}
\centering
\includegraphics[angle=0, width=0.5\textwidth]{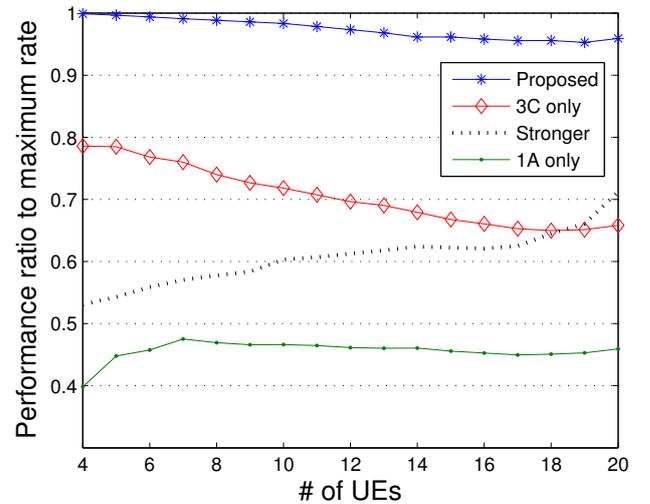}
\caption{Performance Ratio of The Proposed algorithm to the
Optimal Solution} \label{fig:main}
\end{figure}

\begin{figure}
\centering
\includegraphics[angle=0, width=0.5\textwidth]{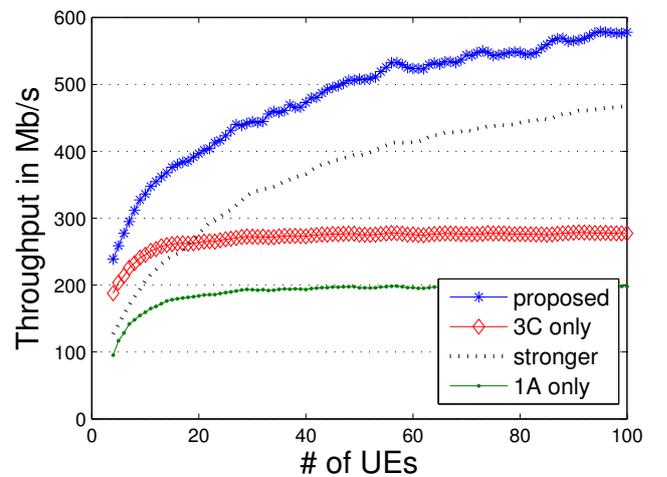}
\caption{Performance comparison with other algorithms}
\label{fig:main_B}
\end{figure}

We show first, the proposed algorithm's performance by
comparison to the optimal solution's capacity as depicted at
Fig. \ref{fig:main}. It shows that the numerical results of
the proposed algorithm are comparable to the optimal solution
within 5 \% gap. On the other hand, applying single profile
algorithm such as `3C only' and `1A only' do not increase
performance as the number of UEs increases because higher UE
density do not make UE's SINR better on average.  In case of
`Stronger' algorithm, though it increases network capacity,
there is still considerable difference to the optimal solution.

Fig. \ref{fig:main_B} shows the network capacities for each
algorithms except the optimal solution, due to the enormous
computation cost of it. The proposed algorithm makes the
network capacity increase as the number of UEs increase. It
comes from that the proposed algorithm provide more frequency
resource to better SINR UE by increasing opportunity for
allocating both MBS and SBS. Despite the `Stronger' algorithm
also make the network capacity increase, there is around 30 \%
gap to the proposed algorithm.

\begin{figure}
\centering
\includegraphics[angle=0, width=0.5\textwidth]{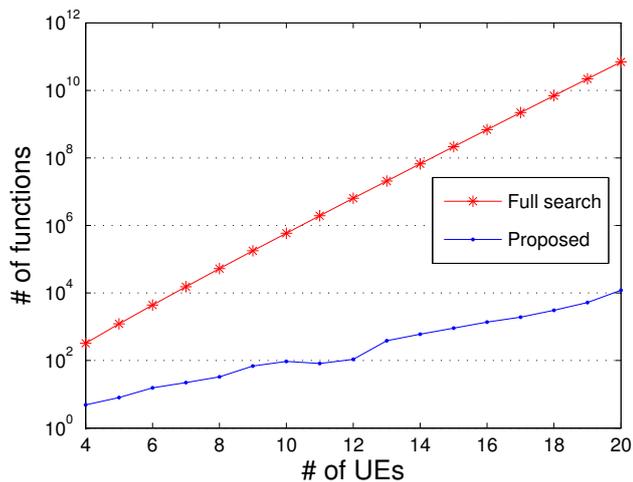}
\caption{Complexity comparison with full search algorithms}
\label{fig:complexity}
\end{figure}

The comparison result for the complexity to the full search
algorithm is shown in Fig. \ref{fig:complexity}. We analyze
the computational complexity of the algorithms by counting
the functions of rate calculation in \eqref{Eq:rate_m} and \eqref{Eq:rate_s}. 
Although both complexities of each
algorithm increase exponentially as the number of UEs
increases, the proposed algorithm's complexity does not exceed
$10^4$ in most cases with the number of UEs not exceeding 20
UEs. Note that the complexity gap between full search and the proposed algorithm increases exponentially so that it is more
than $10^6$ times for a number of UEs greater than 16. At that
time, the proposed algorithm sustains its high performance.

\section{Conclusion}

We have investigated the DC profile allocation problem to
maximize the network capacity in HCNs. 
Although an optimal
algorithm achieves maximum sum rate capacity, it is hard to
apply it because of its exorbitant complexity. In this paper,
we introduced a sub-optimal algorithm that makes network
capacity close to maximum sum rate for a few UEs
(not greater than 20), but considerably lower complexity. 
The simulation result reveals that the proposed algorithm achieves
97\% of the maximum capacity and low complexity up to $10^{-6}$
times compared with the full search algorithm in terms of the number of executed rate-calculation operations.

\end{document}